\documentclass[preprint,aps,prd,10pt,superscriptaddress,twocolumn,showpacs]{revtex4-1}
\usepackage{hyperref}
\usepackage{graphicx}
\usepackage{amsmath,amssymb}
\usepackage{subfigure,dcolumn}
\usepackage[T2A,T1]{fontenc}
\usepackage[russian,english]{babel}
\usepackage{epstopdf}
\usepackage[dvipsnames]{xcolor}
\usepackage{color}
\usepackage{multirow,booktabs}
\usepackage{lipsum}
\usepackage{listings}
\newcommand{\im}{\textrm{i}}
\newcommand{\dint}{\textrm{d}}

\newcommand{\Tab}{Table}
\newcommand{\tab}[1]{\Tab~\ref{tab:#1}}
\begin{document}

\title{Non-standard interactions versus planet-scale neutrino oscillations}

\author{Wei-Jie Feng}
 \email{fengwj7@mail2.sysu.edu.cn}
\author{Jian Tang}%
 \email{tangjian5@mail.sysu.edu.cn}
\author{Tse-Chun Wang}%
 \email{wangzejun@mail.sysu.edu.cn}
\author{Yi-Xing Zhou}%
 \email{zhouyx45@mail2.sysu.edu.cn}
\affiliation{ 
School of Physics, Sun Yat-Sen University, Guangzhou 510275, China
}%

\begin{abstract}
	The low-energy threshold and the large detector size of Precision IceCube Next Generation Upgrade (PINGU) can make the study on neutrino oscillations with a planet-scale baseline possible. In this task, we consider the configuration that neutrinos are produced at CERN and detected in the PINGU detector, as a benchmark. We discuss its sensitivity of measuring the size of non-standard interactions (NSIs) in matter, which can be described by the parameter $\epsilon_{\alpha\beta}$ ($\alpha$ and $\beta$ are flavors of neutrinos). We find that the CERN-PINGU configuration improves $\tilde{\epsilon}_{\mu\mu}\equiv\epsilon_{\mu\mu}-\epsilon_{\tau\tau}$ and $\epsilon_{\mu\tau}$ significantly compared to the next-generation accelerator neutrino experiments. Most of degeneracy problems in the precision measurements can be resolved, except the one for $\tilde{\epsilon}_{\mu\mu}\sim-0.035$. Moreover, we point out that this configuration can also be used to detect the CP violation brought by NSIs. Finally, we compare the physics potential in this configuration to that for DUNE, T2HK and P2O, and find that the CERN-PINGU configuration can significantly improve the sensitivity to NSIs.
\end{abstract}


\maketitle

\section{Introduction}
Since confirming this phenomenon of neutrino oscillations in $1998$~\cite{Kajita:1998bw}, we nearly complete the knowledge of this flavour-changing behaviour, which can be described by six oscillation parameters including three mixing angles $ \theta_{12} $, $ \theta_{13} $ and $ \theta_{23} $, two mass-square differences $ \Delta m_{21}^2 $and $\Delta m_{31}^2$, and one Dirac CP violating phase $ \delta$ with solar, atmospheric, accelerator and reactor neutrino data~\cite{Esteban:2018azc,deSalas:2018bym,Capozzi:2018ubv}. The rest of problems in neutrino oscillations are if $\theta_{23}$ is larger or smaller than $45^\circ$,  which the sign of $\Delta m_{31}^2$ is, if CP is violated and what its value $\delta$ is. These problems are expected to be resolved in the next-generation neutrino oscillation experiments, \textit{e.g.} DUNE, T2HK, JUNO, etc. The neutrino oscillation reflects the fact that neutrinos are massive, which conflicts with the massless-neutrino prediction in the standard model (SM). This phenomenon is obviously a key to the door of physics beyond SM (BSM), and reveals that SM is not a complete theory.

Far away in Antarctica, Precision IceCube Next Generation Upgrade (PINGU), an extension of the IceCube Neutrino Observatory, was proposed. In this proposal, the lower energy threshold and the $6$-million-ton detector are sketched~\cite{TheIceCube-Gen2:2016cap}. Goals of PINGU are to detect atmospheric neutrinos~\cite{choubey2014bounds,winter2013neutrino}, supernova neutrinos~\cite{pingu2014letter} and the indirect signal of dark matter (by detecting the self annihilation of WIMP-like dark matter~\cite{koskinen2011icecube}). Moreover, the configuration that PINGU receives neutrino beams from accelerators in the northern hemisphere inspired by CERN to Frejus and J-PARC to HyperKamiokande has been considered and discussed~\cite{Tang2012}. 

As we believe that SM is not a complete theory, effects from some exotic interactions are widely discussed, for example, non-standard interactions (NSIs)~\cite{Ohlsson:2012kf,Miranda:2015dra,Farzan:2017xzy,Esteban:2018ppq,Dev:2019anc}, neutrino decays~\cite{Acker:1991ej,Acker:1993sz,Tang:2018rer}, nonunitarity~\cite{PhysRevD.22.2227,Miranda:2018yym}. NSIs are interactions evolving at least one neutrinos and other SM fermions by mediating BSM particles~\cite{wolfenstein1978neutrino,Guzzo:1991hi}. NSIs may take place in three different parts of neutrino oscillations: at the source, at the detector, and in matter (or NSI matter effects). We describe the size of NSIs by the parameter $\epsilon^{ff'}_{\alpha\beta}$, which is the fraction of the strength of coupling for the NSI to the Fermi constant ($\nu_\alpha+f\rightarrow\nu_\beta+f'$). For those taking place in matter, we use the notation $\epsilon_{\alpha\beta}$, as in these interactions $f=f'$. 
In recent studies, the Large-Mixing-Angle dark solution (LMA-Dark solution) for NSIs allows that NSI matter effects have a strong impact on neutrino oscillations. This solution predicts $ \epsilon_{ee} \simeq -3 $~\cite{Miranda:2015dra,Esteban:2018ppq}. %mark
One of upcoming long baseline accelerator neutrino experiments Deep Underground Neutrino Experiment (DUNE) \cite{Acciarri:2015uup,Abi:2018dnh} is expected to have some sensitivity to NSIs in matter. Impacts of NSI matter effects on the precision measurement of oscillation parameters and the expected constraints on NSI parameters for DUNE are widely studied~\cite{Coloma:2015kiu,Masud:2015xva,Liao:2016hsa,Agarwalla:2016fkh,Masud:2016gcl,Forero:2016cmb,Masud:2016bvp,Ge:2016dlx,Huitu:2016bmb,Ghosh:2017ged,Flores:2018kwk,Masud:2018pig,Capozzi:2019iqn,Ghoshal:2019pab}. The combined results for $\sin^22\theta_{13}$ and NSIs by Daya Bay and T2K experiments were given in Ref.~\cite{Girardi:2014kca}. The current global fits can provide constraints at $95\%$ C.L. on NSI parameters with COHERENT data~\cite{Esteban:2018ppq}:
	\begin{align*}
	-0.008 \leqslant \varepsilon_{ee}^u \leqslant 0.618\\
	-0.111 \leqslant \varepsilon_{\mu\mu}^u \leqslant 0.402\\
	-0.110 \leqslant \varepsilon_{\tau\tau}^u \leqslant 0.404\\
	-0.006 \leqslant \varepsilon _{e\mu}^u \leqslant 0.049\\
	-0.248 \leqslant \varepsilon_{e\tau}^u \leqslant 0.116\\
	-0.012 \leqslant \varepsilon_{\mu\tau}^u \leqslant 0.009.
	\end{align*}
	Some possible theoretical models are also proposed to realize sizeable NSIs in matter~\cite{Farzan:2015doa,Farzan:2015hkd,Farzan:2016wym,Forero:2016ghr,Babu:2019mfe}.

NSIs in matter can be detected by accelerator neutrino oscillation experiments with non-negligible matter effects. In this paper, we study the planet-scale neutrino oscillations to measure the size of NSI matter effects, by revisiting the configuration of sending a neutrino beam from an accelerator facility in the northern hemisphere such as CERN to a detector in the south pole like PINGU. The CERN-PINGU configuration has a baseline of $11810$~km. The difference of this configuration from observatories of atmospheric neutrinos is about neutrino fluxes, which come from a specific direction with the well-controlled timing structure in the CERN-PINGU configuration. As a result, smaller systematic errors and the higher ratio of signals over backgrounds are expected. This super long baseline has four advantages for $\epsilon_{\alpha\beta}$ measurements as follows.
\begin{enumerate}
	\item The neutrino energy ($3$~GeV - $20$~GeV) is higher than current and future neutrino oscillation experiments. Therefore, larger matter effects are expected, and the detection of $\nu_\tau$ and $\bar{\nu}_\tau$ is achievable. 
	\item As neutrinos propagate in a longer baseline, NSI matter effects are expected to be more important. With the help of this 11810 km baseline, the value of NSI parameters can be measured with higher accuracy. 
	\item If neutrinos propagate through the core with a larger matter density, the effect of NSI in matter will be greater. As a result, the matter density, which can reach $11$~g/cm$^3$, makes this configuration special and promising for the measurement of $\epsilon_{\alpha\beta}$.
	\item It is necessary to mention that though the statistics is lower as the baseline is much longer, this drawback can be compensated by the million-ton detector as PINGU. 
\end{enumerate}

This paper is arranged as follows. We will firstly demonstrate the neutrino oscillation probability with NSI matter effects, before presenting the simulation details. In the following, we will show the constraints on $\epsilon_{\alpha\beta}$ and constraint contours between any two of NSI parameters for the CERN-PINGU configuration. Moreover, we will discuss how this configuration can exclude the CP conserved scenario once the phase of $ \epsilon_{\alpha\beta} $ is non-zero ($ \alpha\ne\beta $). Finally, we will summarize our results, and provide our conclusion.

\section{Neutrino Oscillation Physics with NSIs}
In this section, we briefly introduce how neutrino oscillation probabilities are modified by NSIs in matter. These new interactions are neutral-current-like interactions $\nu_\alpha+f\rightarrow\nu_\beta+f$, and can be described by the operator,  \cite{wolfenstein1978neutrino,GROSSMAN1995141,BEREZHIANI2002207},: 
\begin{equation}\label{equ:LNSI}
\mathcal{L}_{\mathrm{NSI}}=-2 \sqrt{2} G_{F} \varepsilon_{\alpha \beta}^{f}\left(\overline{\nu_{\alpha}} \gamma^{\mu} P_{L} \nu_{\beta}\right)\left(\overline{f} \gamma_{\mu} P_{C} f^{\prime}\right)
\end{equation}
where $G_F$ is the Fermi constant. We note that the total size of NSI matter effects is the sum of those for NSIs with electrons, neutrons and protons: $\epsilon_{\alpha\beta}\equiv \epsilon^e_{\alpha\beta}+\epsilon^n_{\alpha\beta}+\epsilon^p_{\alpha\beta}$. 

Neutrino oscillations are governed by coherent evolution of quantum states
\begin{equation}\label{equ:shro}
\im\frac{\dint}{\dint t}
\begin{bmatrix}
\nu_e \\ \nu_\mu \\ \nu_\tau
\end{bmatrix}
=H
\begin{bmatrix}
\nu_e \\ \nu_\mu \\ \nu_\tau
\end{bmatrix}.
\end{equation}
Explicitly, the evolution of the neutrino flavor state is determined by the Hamiltonian:
\begin{equation}\label{equ:ham}
\begin{array}{l}
	H_\nu = H_{\mathrm{vac}} + H_{\mathrm{mat}}~~~~~~~~~\text{for}~\nu,\\
	 H_{\bar{\nu}}= [H_{\mathrm{vac}} - H_{\mathrm{mat}}]^*~~~~~~\text{for}~\bar{\nu},
\end{array}
\end{equation}
where $ H_{\mathrm{vac}} $ is the Hamiltonian in vacuum. 
$ H_{\mathrm{mat}} $ corresponds to the matter, and is written by
\begin{equation}\label{equ:Hmat}
	H_{\mathrm{mat}} = \sqrt{2}G_FN_e
	\begin{bmatrix}
	1+\epsilon_{ee} &\epsilon_{e\mu} &\epsilon_{e\tau}\\
	\epsilon^{*}_{e\mu} & \epsilon_{\mu\mu} & \epsilon_{\mu\tau}\\
	\epsilon^{*}_{e\tau} & \epsilon^{*}_{\mu\tau} & \epsilon_{\tau\tau}\\
	\end{bmatrix}
\end{equation}
where $ N_e $ is the number density of electron. The constant term $\sqrt{2}G_F N_e$ in the $ ee $ component refers to the standard matter effect~\cite{wolfenstein1978neutrino}. It is obvious in Eqs.~(\ref{equ:shro}) and (\ref{equ:Hmat}) that one of the diagonal terms can be absorbed by an overall phase in neutrino states. Therefore, we define $\tilde{\epsilon}_{ee}\equiv\epsilon_{ee}-\epsilon_{\tau\tau}$ and $\tilde{\epsilon}_{\mu\mu}\equiv\epsilon_{\mu\mu}-\epsilon_{\tau\tau}$ without a loss of generality.

{\subsection{Analytical approximation}\label{Sec:analytical_prob}

The main probabilities for the CERN-PINGU configuration to measure the effect of NSIs are via $P(\nu_\mu\rightarrow\nu_\mu)$ and $P(\nu_\mu\rightarrow\nu_e)$ and their CP-conjugate channels, because of the difficulty of the $\nu_\tau$ and $\bar{\nu}_\tau$ detection. 
Taking $\frac{\Delta m_{21}^2}{\Delta m_{31}^2}\sim |\epsilon_{\alpha\beta}|\sim s^2_{13}$ as the first order of perturbation $\xi$, the approximation equations for the probability,
\begin{align}\label{equ:Pmumu}
P(\nu_\mu\rightarrow\nu_\mu)&=P_0(\nu_\mu\rightarrow\nu_\mu)+\delta P_{\text{NSI}} (\nu_\mu\rightarrow\nu_\mu) \nonumber\\ 
&= P_0(\nu_\mu\rightarrow\nu_\mu)-A\epsilon_{\mu\tau}\cos\phi_{\mu\tau} \nonumber\\ 
& \times\left(\sin^32\theta_{23}\frac{L}{2E}\sin2\Delta_{31}\right. \nonumber\\ 
&\left.+4\sin2\theta_{23}\cos^22\theta_{23}\frac{1}{\Delta m_{31}^2}\sin^2\Delta_{31}\right)\nonumber\\ 
& +A\tilde{\epsilon}_{\mu\mu}c^2_{23}s^2_{23}\left(c_{23}^2-s^2_{23}\right)\nonumber\\ 
& \times\left(\frac{L}{E}\sin2\Delta_{31}-\frac{8}{\Delta m_{31}^2}\sin^2\Delta_{31}\right)\nonumber\\ 
& +\mathcal{O}(\xi^2),
\end{align}

\begin{align}\label{equ:Pmue}
P(\nu_\mu\rightarrow\nu_e)&=P_0(\nu_\mu\rightarrow\nu_e)+\delta P_{\text{NSI}}(\nu_\mu\rightarrow\nu_e)\nonumber\\ 
& =P_0(\nu_\mu\rightarrow\nu_e)+8s_{13}|\epsilon_{e\mu}|s_{23}\frac{\Delta m_{31}^2}{\Delta m_{31}^2-A}\sin\Delta_{31}^A\nonumber\\  
& \times\left(s_{23}^2\frac{A}{\Delta m_{31}^2-A}\cos(\delta+\phi_{e\mu})\sin\Delta^A_{31}\right.\nonumber\\ 
& \left.+c^2_{23}\sin\frac{AL}{4E}\cos(\delta+\phi_{e\mu}-\Delta_{31})\right)\nonumber\\ 
& +8s_{13}|\epsilon_{e\tau}|c_{23}s_{23}^2\frac{\Delta m_{31}^2}{\Delta m_{31}^2-A}\sin\Delta^A_{31}\nonumber\\ 
& \times\left(\frac{A}{\Delta m_{31}^2-A}\cos(\delta+\phi_{e\tau})\sin\Delta^A_{31}\right. \nonumber\\ 
& \left.-\sin\frac{AL}{4E}\cos(\delta+\phi_{e\tau}-\Delta_{31})\right)\nonumber\\ 
& +\mathcal{O}(\xi^2),
\end{align}
where $P_0(\nu_\mu\rightarrow\nu_\mu)$ and $P_0(\nu_\mu\rightarrow\nu_e)$ are the probability for $\nu_\mu\rightarrow\nu_\mu$ and $\nu_\mu\rightarrow\nu_e$ channels without NSIs, respectively. And the notations $\Delta_{31}$ and $\Delta^A_{31}$ are defined as 
\begin{equation}
\Delta_{31}\equiv\frac{\Delta m_{31}^2L}{4E},~\text{and}~\Delta m_{31}^A\equiv\frac{\Delta m_{31}^2-A}{4E}\times L, 
\end{equation}
with $A\equiv 2\sqrt{2}G_FN_eE$. For antineutrino modes, the factors $A$ and $\delta$ are replaced by $-A$ and $-\delta$, respectively. These equations are consistent with Ref.~\cite{Wang:2018dwk,Kikuchi:2008vq}. The impact of NSIs is proportional to $A$. As a result, the increase of matter density enhances the measurement capacity of NSIs. In the configuration of CERN-PINGU, the matter density can reach up to $11$ g/cm$^3$, which is roughly three-time larger than the average matte density for DUNE ($\sim3$ g/cm$^3$). Moreover, we notice that in $P(\nu_\mu\rightarrow\nu_\mu)$, one term for $\epsilon_{\mu\tau}$ and $\tilde{\epsilon}_{\mu\mu}$ are proportional to $L$, of which for the CERN-PINGU configuration is about nine-time longer than the baseline for DUNE $1300$~km. We can expect the larger improvement in the measurement for these two NSI parameters from DUNE by the CERN-PINGU configuration. %And, this is consistent with what we found in Fig.~\ref{fig:prob}. 
The same dependence on $L$ is also seen for $\epsilon_{e\mu}$ and $\epsilon_{e\tau}$ in the appearance channel. However, they are higher order terms than $\tilde{\epsilon}_{\mu\mu}$ and $\epsilon_{\mu\tau}$. 

The impact of $\phi_{e\mu}$ or $\phi_{e\tau}$ on $P(\nu_\mu\rightarrow\nu_e)$ depends on the value of $\delta$ and $\Delta_{31}$. This dependence is not seen for $\phi_{\mu\tau}$ in $P(\nu_\mu\rightarrow\nu_\mu)$. As a result, the impact of $\delta$ and $\Delta m^2_{31}$ is larger in the measurement of $\phi_{e\mu}$ or $\phi_{e\tau}$. Also, $\epsilon_{e\mu}$ and $\epsilon_{e\tau}$ are the higher order than $\epsilon_{\mu\tau}$. Therefore, we can expect it is easier to detect the CP violation by $\phi_{\mu\tau}$, and even measure its size. 

To sum up, we expect that the NSI measurement by the CERN-PINGU configuration can be better than what DUNE can achieve, because of the larger matter density. Though the matter density for this configuration is not overwhelmingly larger than that for DUNE. But the nine-time longer baseline can largely improve the measurement for $\epsilon_{e\tau}$ and $\tilde{\epsilon}_{\mu\mu}$ by the disappearance channels.

\subsection{Probabilities by numerical calculations}\label{sec:numerical_P}
}

\begin{figure}[!ht]
	\centering
	\includegraphics[width=0.45\linewidth]{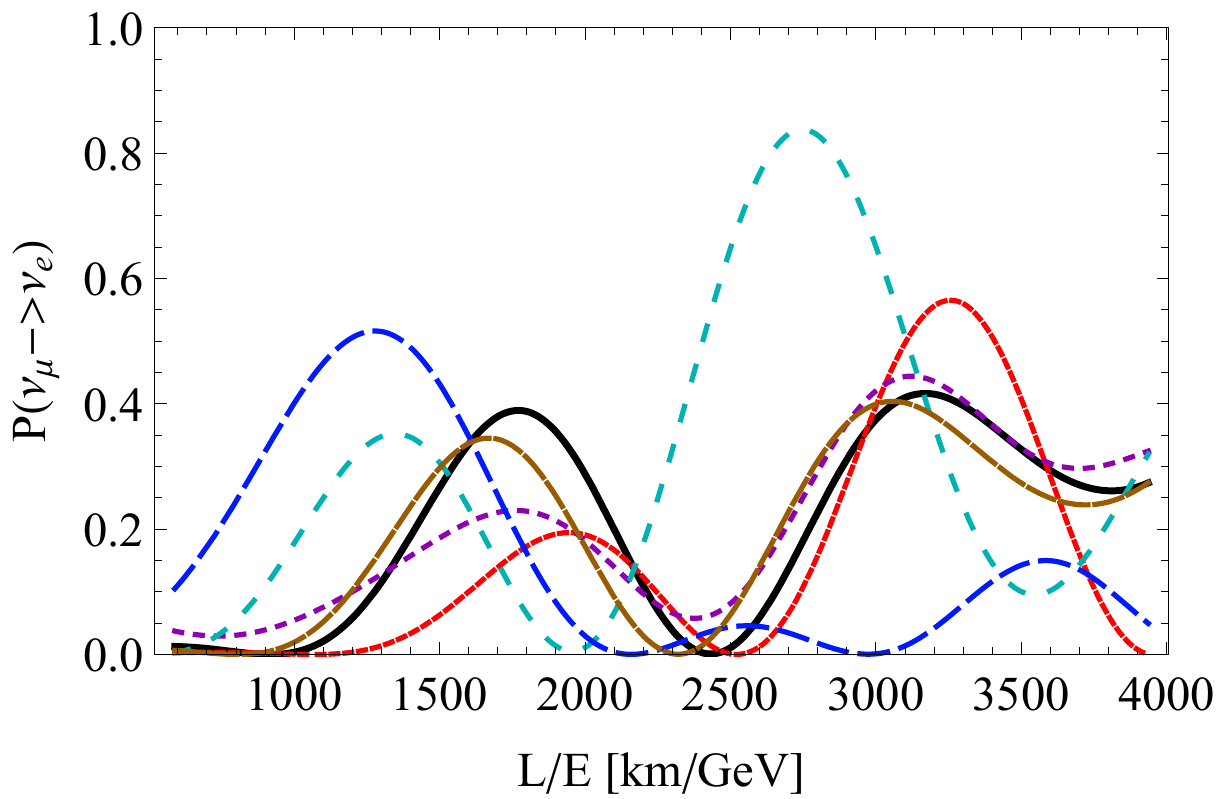}
	\includegraphics[width=0.45\linewidth]{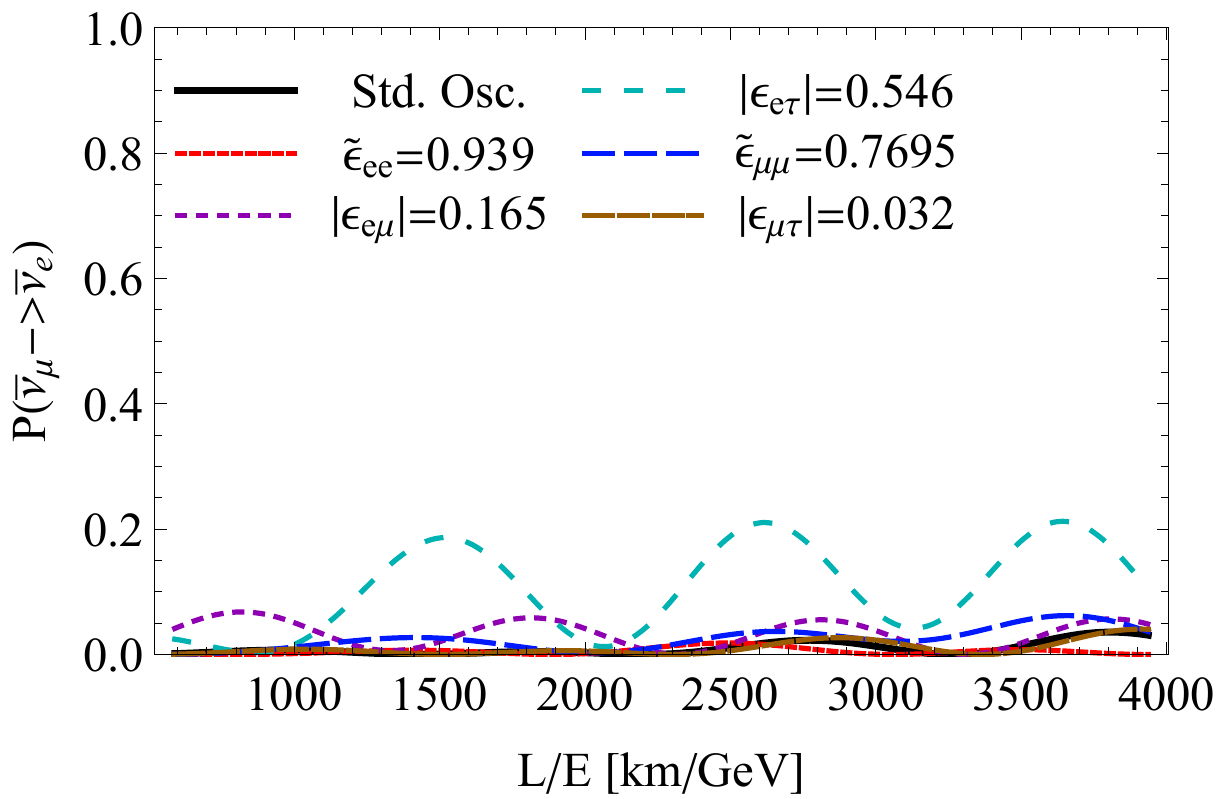}\\
	\includegraphics[width=0.45\linewidth]{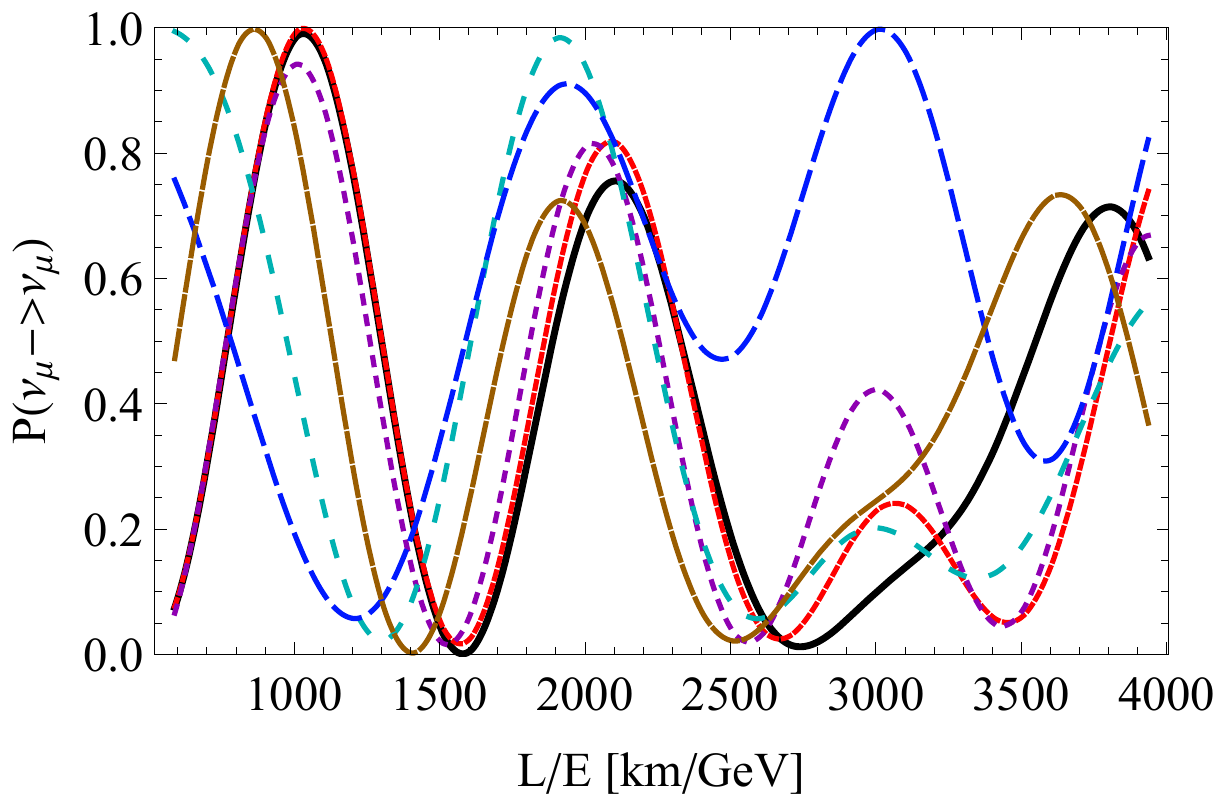}
	\includegraphics[width=0.45\linewidth]{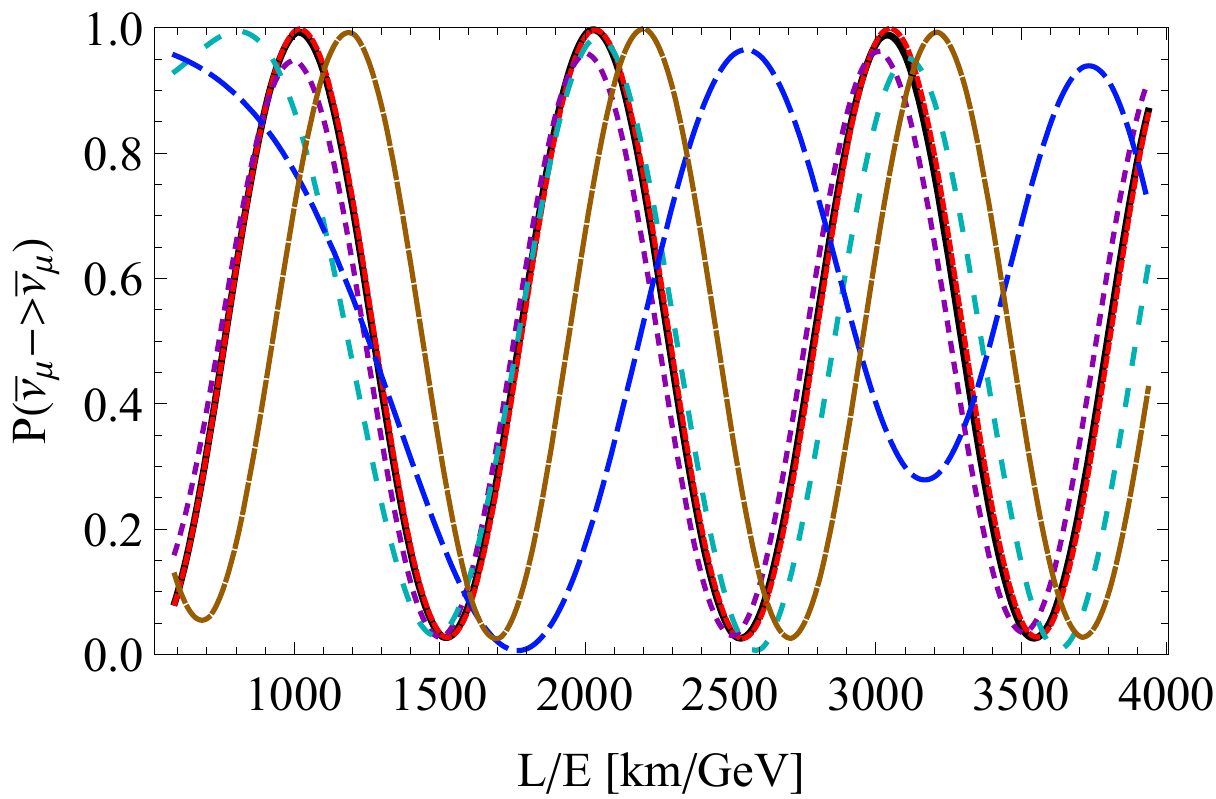}\\
	\includegraphics[width=0.45\linewidth]{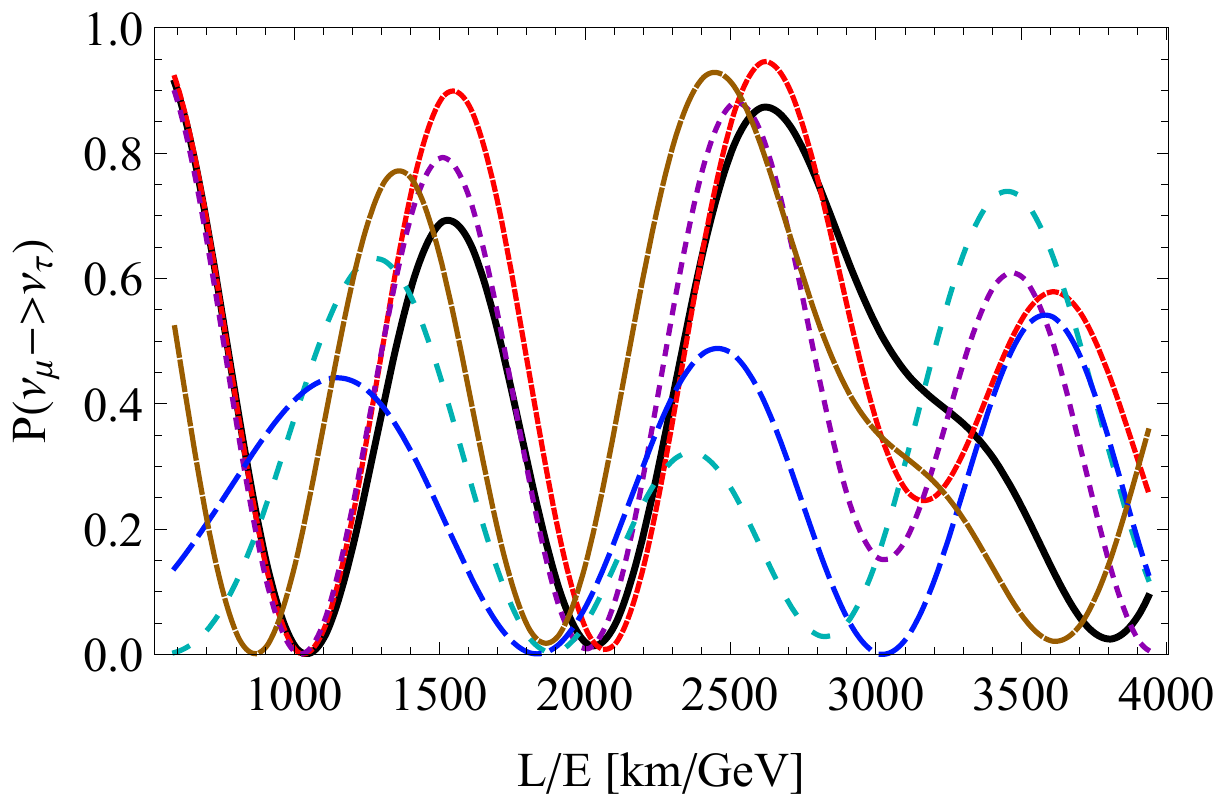}
	\includegraphics[width=0.45\linewidth]{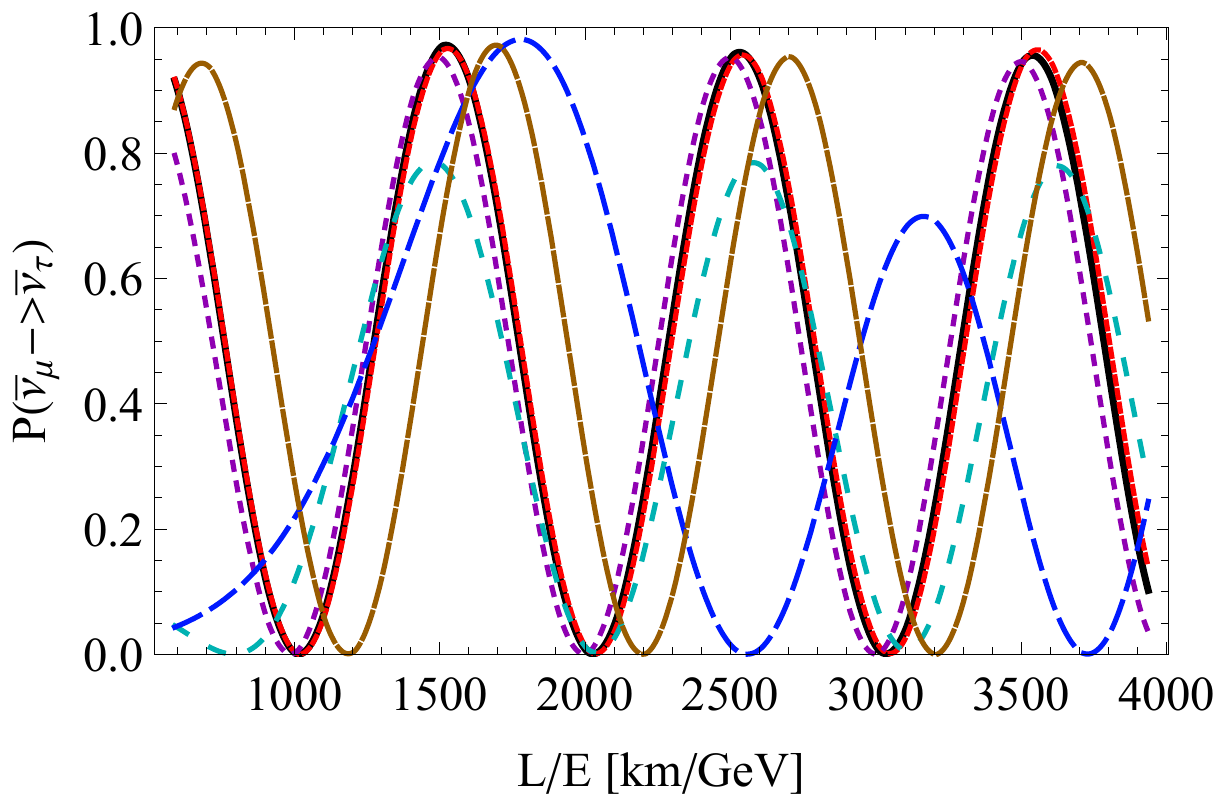}\\
	\caption{The neutrino oscillation probabilities $P(\nu_\mu\rightarrow\nu_e)$ (upper-left), $P(\bar{\nu}_\mu\rightarrow\bar{\nu}_e)$ (upper-right), $P(\nu_\mu\rightarrow\nu_\mu)$ (central-left), $P(\bar{\nu}_\mu\rightarrow\bar{\nu}_\mu)$ (central-right), $P(\nu_\mu\rightarrow\nu_\tau)$ (lower-left), and $P(\bar{\nu}_\mu\rightarrow\bar{\nu}_\tau)$ (lower-right) in the case with standard matter effects and in the case with nonzeoro $\epsilon_{\alpha\beta}$. The probabilities are shown as functions of $L/E$ [km/GeV] in the range of $3<E/\text{GeV}<20$ with the baseline of $11810$ km. The values are used according to the $1\sigma$ uncertainty of the current global fit result~\cite{Esteban:2018ppq}. All phases are set to be $0$.}
	\label{fig:prob}
\end{figure}

Fig.~\ref{fig:prob} shows probabilities in each of channels in the case only including standard matter effects, and for those with a non-zero $\epsilon_{\alpha\beta}$, of which the value is used according to the $1\sigma$ bound of the current global fit result~\cite{Esteban:2018ppq}. These probabilities are shown as functions of $L/E$ [km/GeV] in the range of neutrino energy ($3$~GeV - $20$~GeV) with the baseline of $11810$ km. 
%We see the behaviour of oscillation probabilities $P(\bar{\nu}_\mu\rightarrow\bar{\nu}_\mu)$ and $P(\bar{\nu}_\mu\rightarrow\bar{\nu}_\tau)$ are similar for all cases. 
%
We obtain that the behavior of probabilities for non-zero $\tilde{\epsilon}_{\mu\mu}$ or $|\epsilon_{\mu\tau}|$ in some channels is very different from that without NSI matter effects. For example, when $L/E=2700$ [km$/$GeV], $P(\nu_\mu\rightarrow\nu_e)$ is about $0.8$ in the case with non-zero $|\epsilon_{\mu\tau}|$, while it is around $0.2$ for the standard case. We also see that the probability of the $\bar{\nu}_\mu$ disappearance channel approaches the maximum at $L/E=2500$ [km$/$GeV] in the non-zero case of $\tilde{\epsilon}_{\mu\mu}$, while this probability is around zero in the framework of standard neutrino oscillations.

\begin{figure}[!ht]
	\centering
	\includegraphics[width=0.45\linewidth]{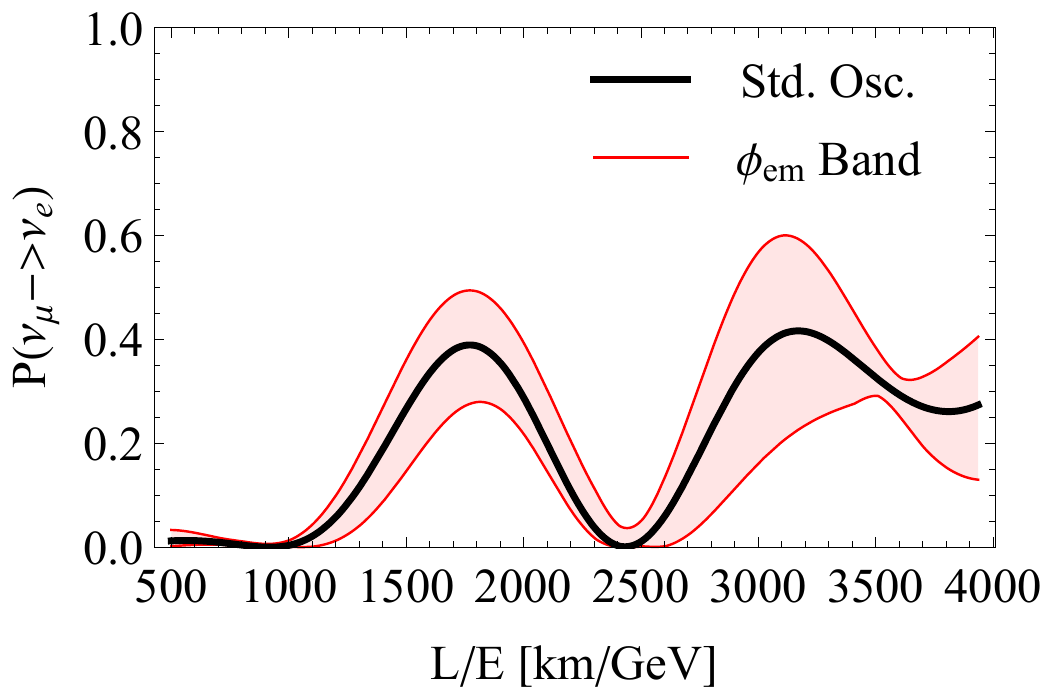}
	\includegraphics[width=0.45\linewidth]{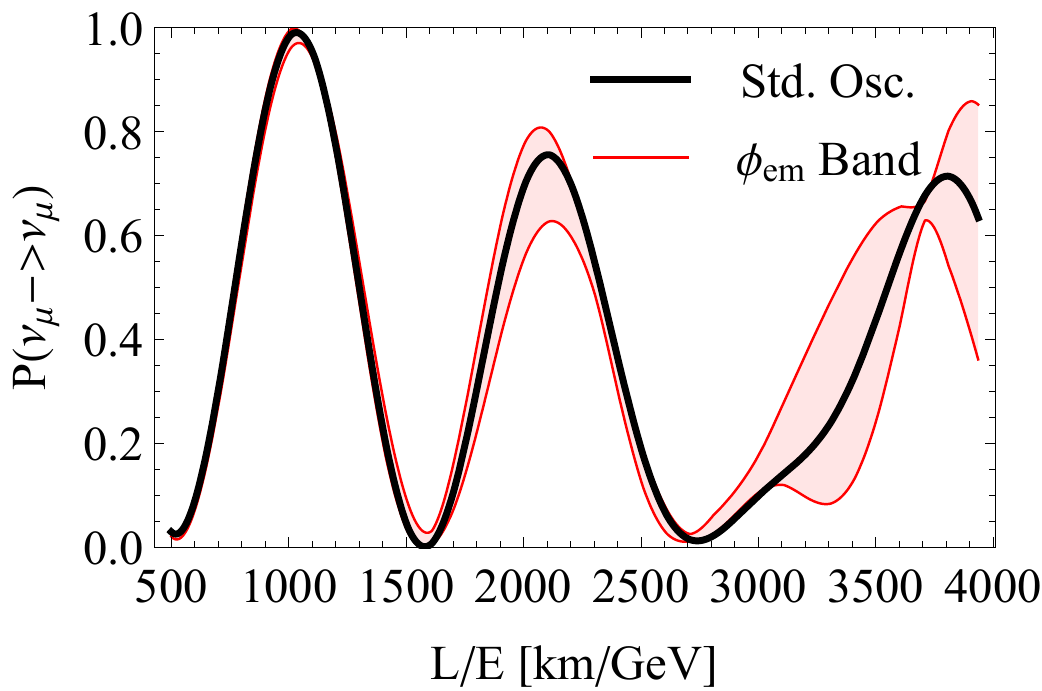}\\
	\includegraphics[width=0.45\linewidth]{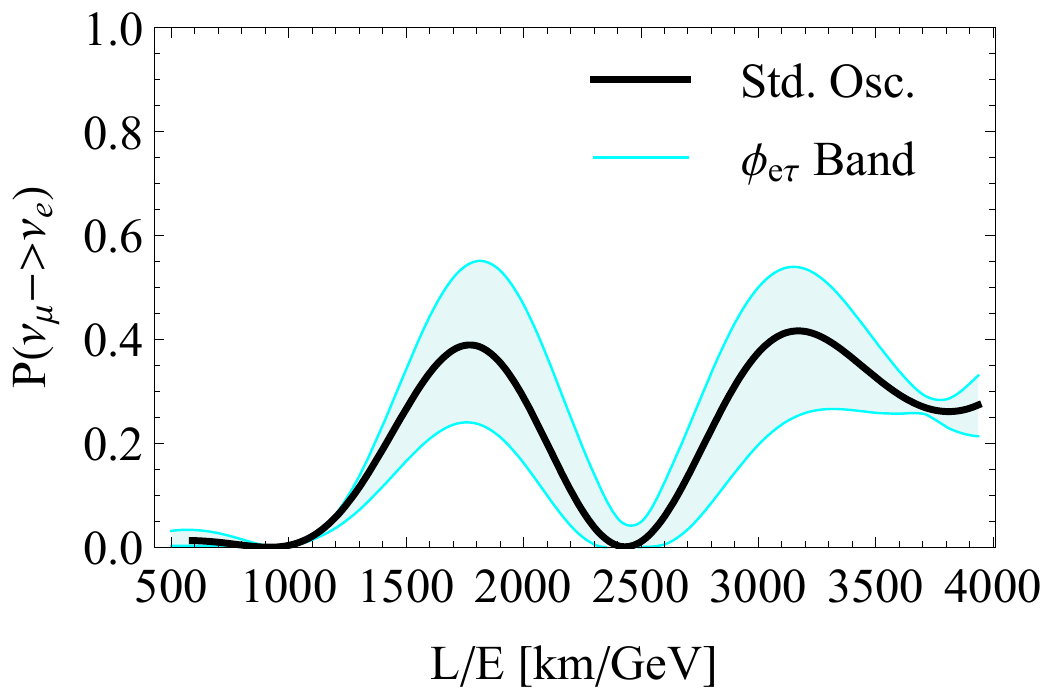}
	\includegraphics[width=0.45\linewidth]{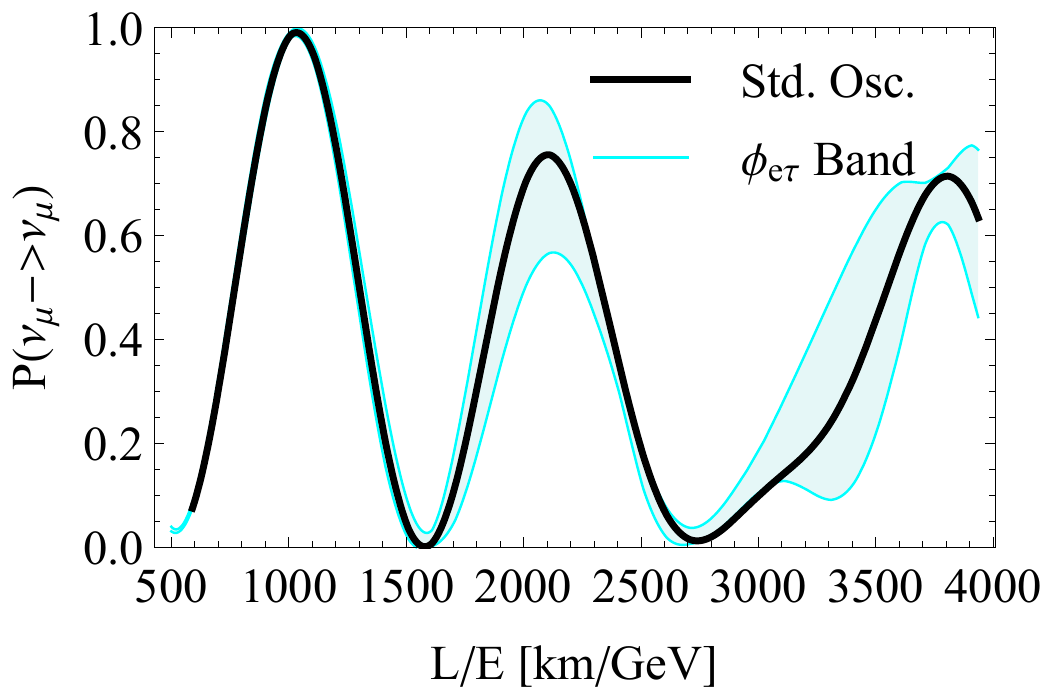}\\
	\includegraphics[width=0.45\linewidth]{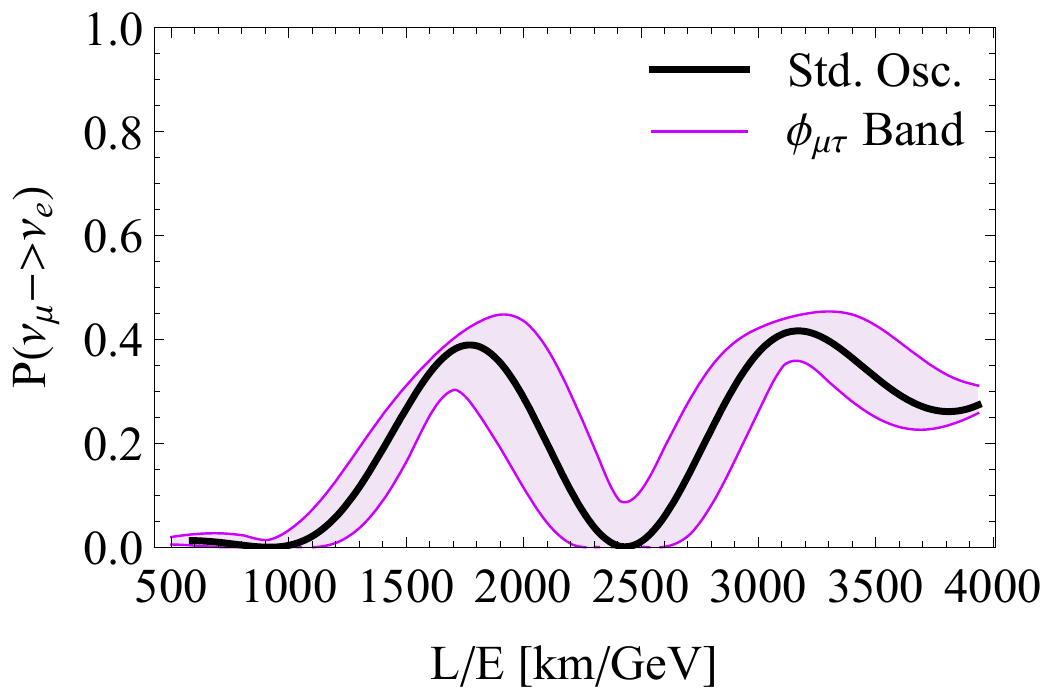}
	\includegraphics[width=0.45\linewidth]{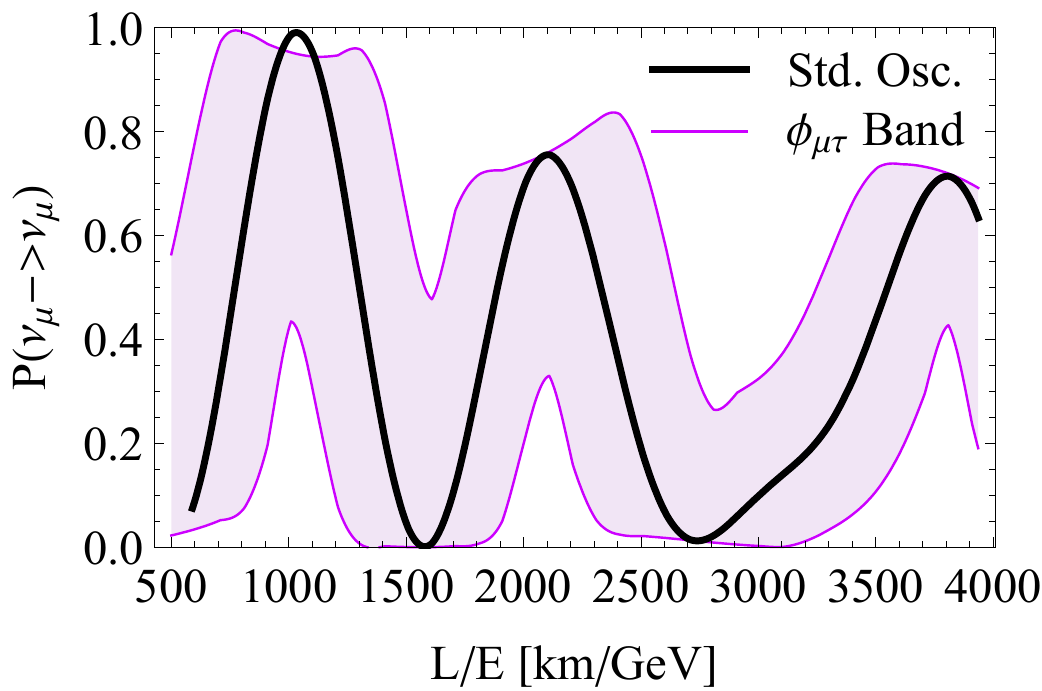}\\
	\caption{The neutrino oscillation probabilities $ P(\nu_\mu\rightarrow\nu_e) $ (left panels), $ P(\nu_\mu\rightarrow\nu_\mu) $ (right panels) for phases $ \phi_{e\mu} $ (pink), $ \phi_{e\tau} $ (light blue) and $ \phi_{\mu\tau} $ (light-magenta) varying over $ (-180^\circ, 180^\circ] $. For the band with non-zero $\phi_{\alpha\beta}$, the absolute value $|\epsilon_{\alpha\beta}|$ is fixed at $0.05$. The probabilities are shown as functions of $L/E$ [km/GeV] in the range of $3<E/\text{GeV}<20$ with the baseline of $11810$ km.}
	\label{fig:prob_phase}
\end{figure}

The probabilities with possible values of $\phi_{\alpha\beta}$ are shown in Fig.~\ref{fig:prob_phase}. We vary each phase $\phi_{\alpha\beta}$ from $-180^\circ$ to $180^\circ$, and fix the value of $|\epsilon_{\alpha\beta}|$ at $0.05$. Two probabilities $P(\nu_\mu\rightarrow\nu_e)$ and $P(\nu_\mu\rightarrow\nu_\mu)$ are shown, because events in the other channels are less than these two.
We find the variation of the probability is the most dramatic for $\phi_{\mu\tau}$. What follows is that for $\phi_{e\tau}$, while the smallest one is for $\phi_{e\mu}$. We see the same result in the other channels.

{
\subsection{The $\theta_{23}$-$\tilde{\epsilon}_{\mu\mu}$ degeneracy}\label{sec:degeneracy}

There is a degeneracy between $\theta_{23}$ and $\tilde{\epsilon}_{\mu\mu}$. This degeneracy can be easier to be understood by further taking $\delta\theta_{23}\equiv\theta_{23}-45^\circ$ as a perturbation at the first order $\mathcal{O}(\xi)$. For the disappearance channel, 
\begin{align}\label{eq:Pmumu_dtheta45}
P(\nu_\mu\rightarrow\nu_\mu)=&P_0(\nu_\mu\rightarrow\nu_\mu)-|\epsilon_{\mu\tau}|\cos\phi_{\mu\tau}\frac{AL}{2E}\sin2\Delta_{31}\nonumber\\
& -2\delta\theta_{23}\tilde{\epsilon}_{\mu\mu} A 
\times\left(\frac{L}{E}\sin2\Delta_{31}\right.
\nonumber\\ & \left.
-\frac{8}{\Delta m_{31}^2}\sin^2\Delta_{31}\right)+\mathcal{O}(\xi^2)
\end{align}
This is obvious that there is a degeneracy between $\delta \theta_{23}$ and $\tilde{\epsilon}_{\mu\mu}$. In our work, $\delta\theta_{23}\sim0.08$ for the true value $\theta_{23}=49.7^\circ$. This degeneracy has also been studied in Ref.~\cite{Coloma:2015kiu,Coloma:2011rq,Kikuchi:2008vq,Friedland:2004ah,Friedland:2006pi}.  %1105.5936, 0809.3312, 0408264, 0606101

The degeneracy between $\tilde{\epsilon}_{ee}$ and $\epsilon_{\tau\tau}$ that is from the higher order terms has been also studied in Ref.~\cite{Coloma:2015kiu,Chaves:2018sih}.

}

\section{Simulation details}
We adopt General Long Baseline Experiment Simulator~\cite{HUBER2005195,HUBER2007432} with the PINGU simulation package. 

For the production, we assume $\mu$ decays driving the production of neutrinos:
\begin{align}\label{equ:sb}
	\pi^-&\rightarrow \mu^- + \nu_{\mu},\\
	\pi^+&\rightarrow \mu^+ + \bar{\nu}_{\mu}.
\end{align}
We set the run time of $5$ years for neutrino and antineutrino modes (total run time is $10$ years) with 1$\times 10^{20}$ protons on target (POTs) per year. The energy of protons is assumed 120 GeV and the power is 708 kW. 

To include NSI matter effects, we adopt the GLoBES extension package mentioned in Refs.~\cite{Kopp:2006wp,PhysRevD.76.013001,PhysRevD.77.013007,PhysRevD.97.035018}. 
{We adopt the PREM onion shell model of the earth for the matter density profile~\cite{stacey_1977, Dziewonski:1981xy}{, which is shown in Appendix~\ref{App:matter_density}.}}
%With this extension, we modify the neutrino oscillation probability engine with NSI matter effects. 
The matter density can reach 11~g$/$cm$^3$. 

\begin{table}[!t]
	\caption{Disappearance and appearance channels considered in this experiment.}
	\centering
	\begin{tabular}{cc}
		\toprule
		Disappearance channels & appearance channels\\
		\midrule
		 \multirow{2}{*}{$ \nu_\mu \rightarrow \nu_\mu $} & $ \nu_\mu \rightarrow \nu_e $\\
		  & $ \bar{\nu}_\mu \rightarrow \bar{\nu}_e $ \\
		 \multirow{2}{*}{$ \bar{\nu}_\mu \rightarrow \bar{\nu}_\mu $} & $ \nu_\mu \rightarrow \nu_\tau $\\
		  & $ \bar{\nu}_\mu \rightarrow \bar{\nu}_\tau $\\
		\bottomrule
	\end{tabular}

	\label{tab:channel}

\end{table}

All oscillation channels are listed in \tab{channel}. We consider the intrinsic $ \nu_e $ and $ \bar{\nu}_e $ backgrounds in the beam and the atmospheric neutrino backgrounds. We note that Water Cherenkov neutrino detector cannot make a distinction between neutrinos and antineutrinos. Therefore, for the electron-flavor channels, we analyse the combination of $\nu_e$ and $\bar{\nu}_e$ spectra. {Because of the difficulty of $\nu_\tau$ and $\bar{\nu}_\tau$ detection, we adopt a relative conserved assumption for detection efficiency for $\nu_\tau$ and $\bar{\nu}_\tau$.}

\begin{figure}[!ht]
	\centering
	\includegraphics[width=0.45\linewidth]{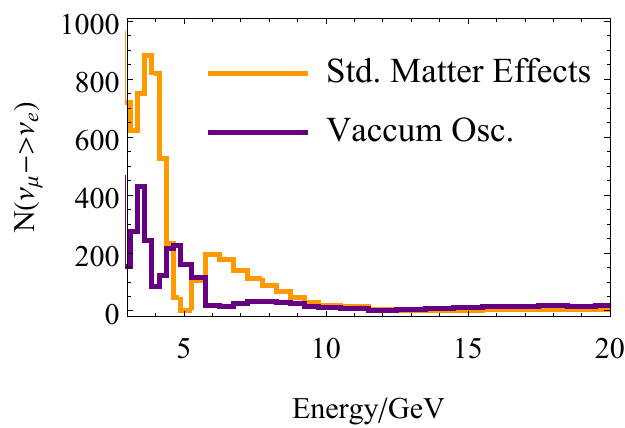}
	\includegraphics[width=0.45\linewidth]{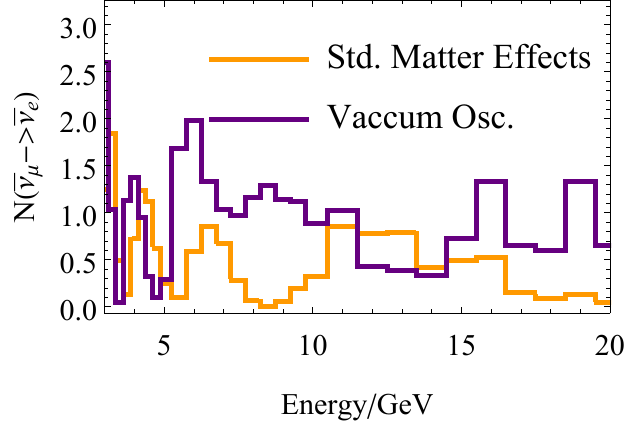}\\
	\includegraphics[width=0.45\linewidth]{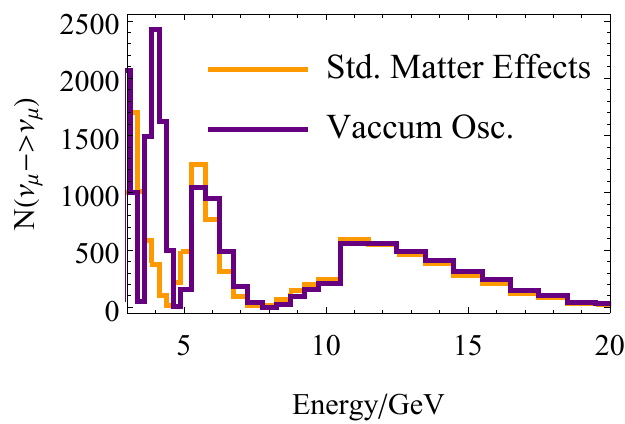}
	\includegraphics[width=0.45\linewidth]{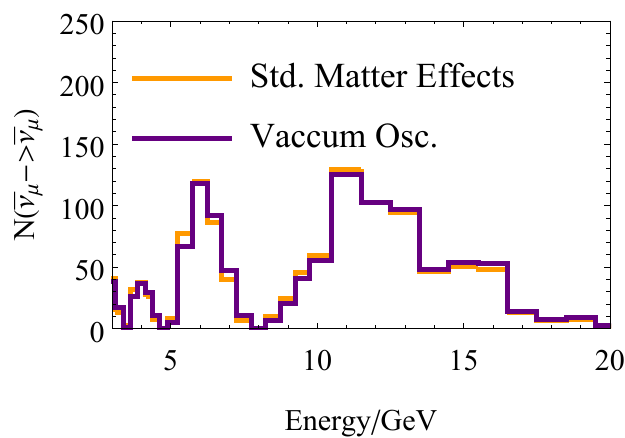}\\
	\includegraphics[width=0.45\linewidth]{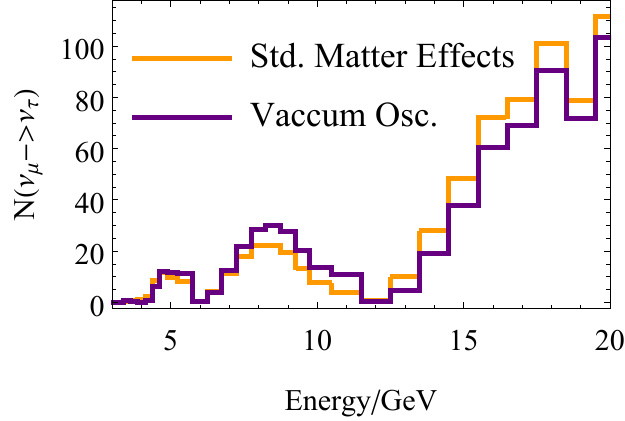}
	\includegraphics[width=0.45\linewidth]{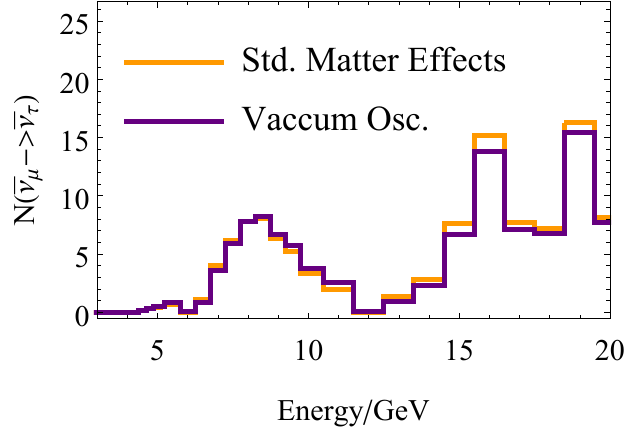}\\
	\caption{The spectra for $P(\nu_\mu\rightarrow\nu_e)$ (upper-left), $P(\bar{\nu}_\mu\rightarrow\bar{\nu}_e)$ (upper-right), $P(\nu_\mu\rightarrow\nu_\mu)$ (central-left), $P(\bar{\nu}_\mu\rightarrow\bar{\nu}_\mu)$ (central-right), $P(\nu_\mu\rightarrow\nu_\tau)$ (lower-left), and $P(\bar{\nu}_\mu\rightarrow\bar{\nu}_\tau)$ (lower-right) with (purple) and without (orange) the standard matter effects. {The $\nu_\tau$ and $\bar{\nu}_\tau$ spectra shown here are optimistic without considering the detection difficulty.}}
	\label{fig:spectra}
\end{figure}

The expected spectra are shown in Fig.~\ref{fig:spectra}. We compare the case with standard matter effects to that in vacuum, for demonstrating the advantage of the CERN-PINGU configuration to collect matter effects through such a long baseline $11810$ km. Assumed the normal mass ordering, events are much less in the $\bar{\nu}$ mode. We see that matter effects make great impacts on the spectra in the $\nu_e$ appearance and $\nu_\mu$ and $\bar{\nu}_\mu$ disappearance channels for the lower neutrino energy. 

\begin{table}[!t]
	\caption{The central value and the width of the prior for standard neutrino oscillation and NSI parameters used throughout our simulation. We note that the central values are as same as the true values, except for the case in which the true values are specified. These values are used according to current global fit results~\cite{Esteban:2018azc,Esteban:2018ppq}.}
	\centering
	\begin{tabular}{ccc}
		\toprule
		parameter & true/central value & 1$\sigma$ width\\
		\midrule
		$\theta_{12}$/$ ^\circ $ & 33.82 & 2\% \\
		$\theta_{13}$/$ ^\circ $ & 8.61 & 2\% \\
		$\theta_{23}$/$ ^\circ $ & 49.7  & 2\%\\
		$ \Delta m_{21}^2 $/$ 10^{-5}\mathrm{eV} $ & 7.38 & 2\% \\
		$ \Delta m_{31}^2 $/$ 10^{-3}\mathrm{eV} $ & 2.525 & 1\% \\
		$ \delta_{\mathrm{CP}} $/$ ^\circ $ & 217 & 10\% \\
		$ \widetilde{\epsilon}_{ee}$ & 0 & 0.939 \\
		$ \widetilde{\epsilon}_{\mu\mu}$ & 0 & 0.7695 \\
		$ |\epsilon_{e\mu}|$ & 0 & 0.165  \\
		$ |\epsilon_{e\tau}|$ & 0 & 0.546 \\
		$ |\epsilon_{\mu\tau}|$ & 0 & 0.0315 \\
		$ \phi_{e\mu}$ & 0 & no restriction\\
		$ \phi_{e\tau}$ & 0 & no restriction  \\
		$ \phi_{\mu\tau}$ & 0 & no restriction  \\

		\bottomrule
	\end{tabular}
	\label{tab:params}
\end{table}

Values used for oscillation and NSI parameters through this paper are listed in \tab{params}. These values are taken from current global-fit results~\cite{Esteban:2018azc,Esteban:2018ppq}. 
The normal mass ordering is assumed. We neglect NSIs at the source and detector, and marginalise over all parameters, including standard oscillation parameters, the matter density and eight parameters for NSIs in matter ($\epsilon_{\tau\tau}$ is subtracted by an overall phase of neutrino states) unless we showed them in the plot.

\section{Results}\label{sec:results}

In this section, our simulation results will be presented. In Sec.~\ref{sec:constraint}, we will firstly show the expected constraints on $\epsilon_{\alpha\beta}$ for the CERN-PINGU configuration. In the following section, we will discuss the sensitivity of CPV in NSIs. In more details, what will be studied further is how much the hypothesis $\phi_{\alpha\beta}=0$ and $\pi$ can be excluded by this configuration, once CP is violated because of non-zero of $\phi_{\alpha\beta}$. 

\subsection{Constraints on NSI parameters}\label{sec:constraint}

\begin{figure}[!ht]
	\centering
	\includegraphics[width=0.49\linewidth]{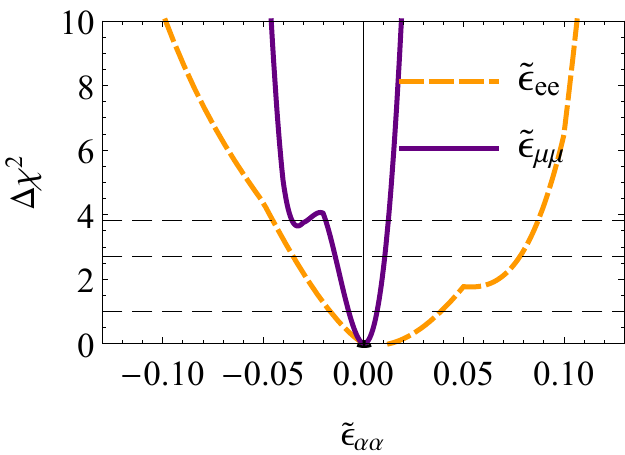}
	\includegraphics[width=0.49\linewidth]{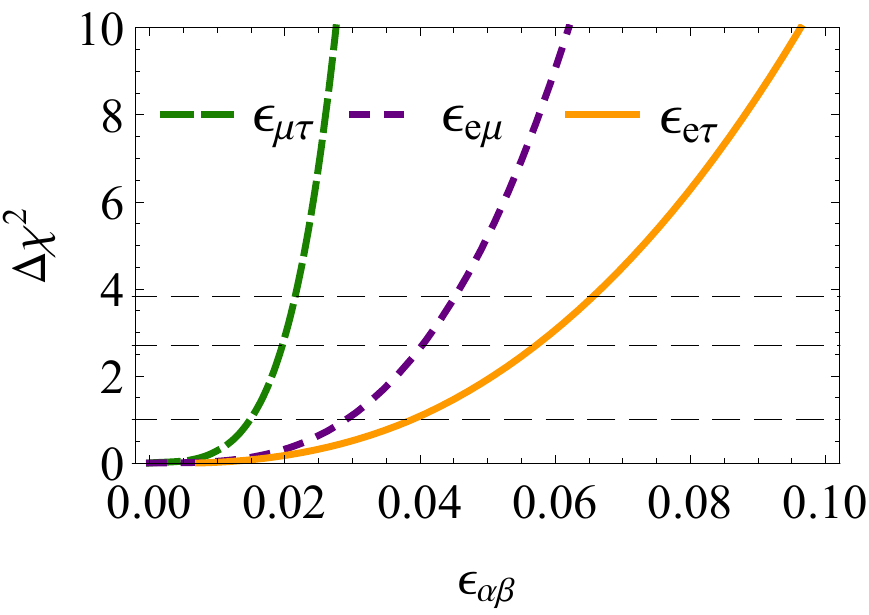}
		\caption{$\Delta\chi^2$ value against $\tilde{\epsilon}_{\alpha\alpha}$ (left) and $|\epsilon_{\alpha\beta}|$ (right). The three dashed lines represent the values at 68\%, 90\% and 95\% C.L. The central values and priors of neutrino mixing parameters are taken from Table~\ref{tab:params}. }
	\label{fig:1D}
\end{figure}

The $\Delta\chi^2$ values against to each NSI parameter are shown in Fig.~\ref{fig:1D}. The size of the $1\sigma$ uncertainty for $\tilde{\epsilon}_{ee}$ is about $0.03$. For the $\mu\mu$ component, the $1\sigma$ error is about $-0.01<\epsilon_{\mu\mu}<0.01$. We find out a degeneracy around $\tilde{\epsilon}_{\mu\mu}=-0.035$ with $\Delta\chi^2\gtrsim4$. For the off-diagonal terms, bounds at  $1\sigma$ C.L. are at $|\epsilon_{e\mu}|\sim0.03$, $|\epsilon_{e\tau}|\sim0.04$, and $|\epsilon_{\mu\tau}|\sim0.015$, respectively. {We reach the conclusion that constraints on $\tilde{\epsilon}_{\mu\mu}$ and $|\epsilon_{\mu\tau}|$ are better than the others, which is consistent with the conclusion of Secs.~\ref{Sec:analytical_prob} and \ref{sec:numerical_P}.}
We review the $1\sigma$ uncertainties in the current global fit result, which are shown in Table~\ref{tab:params}. We see the weak constraint at $1\sigma$ C.L. on the diagonal terms $\tilde{\epsilon}_{ee}$ and $\tilde{\epsilon}_{\mu\mu}$, whose sizes are almost $\mathcal{O}(1)$. As for the off-diagonal terms, we have better understandings in the global fit. The size of $1\sigma$ uncertainty is smaller than $0.1$, especially the $1\sigma$ uncertainty for $\epsilon_{\mu\tau}$ is $0.0315$.
We find that the CERN-PINGU configuration can improve sensitivities of NSIs at $1\sigma$ C.L. by at least $50\%$ of the current global fit result. 
As the uncertainty for $\widetilde{\epsilon}_{ee}$ is greatly improved, we have the interest on the exclusion ability of the LMA-dark solution. Though the result is not shown here, we have checked that the LMA-dark solution can be excluded by more than $ 5\sigma$ C.L., without any prior for $\theta_{12}$ and $\epsilon_{ee}$. 

\begin{figure*}[!bht]
\centering
    \includegraphics[width=0.9\linewidth]{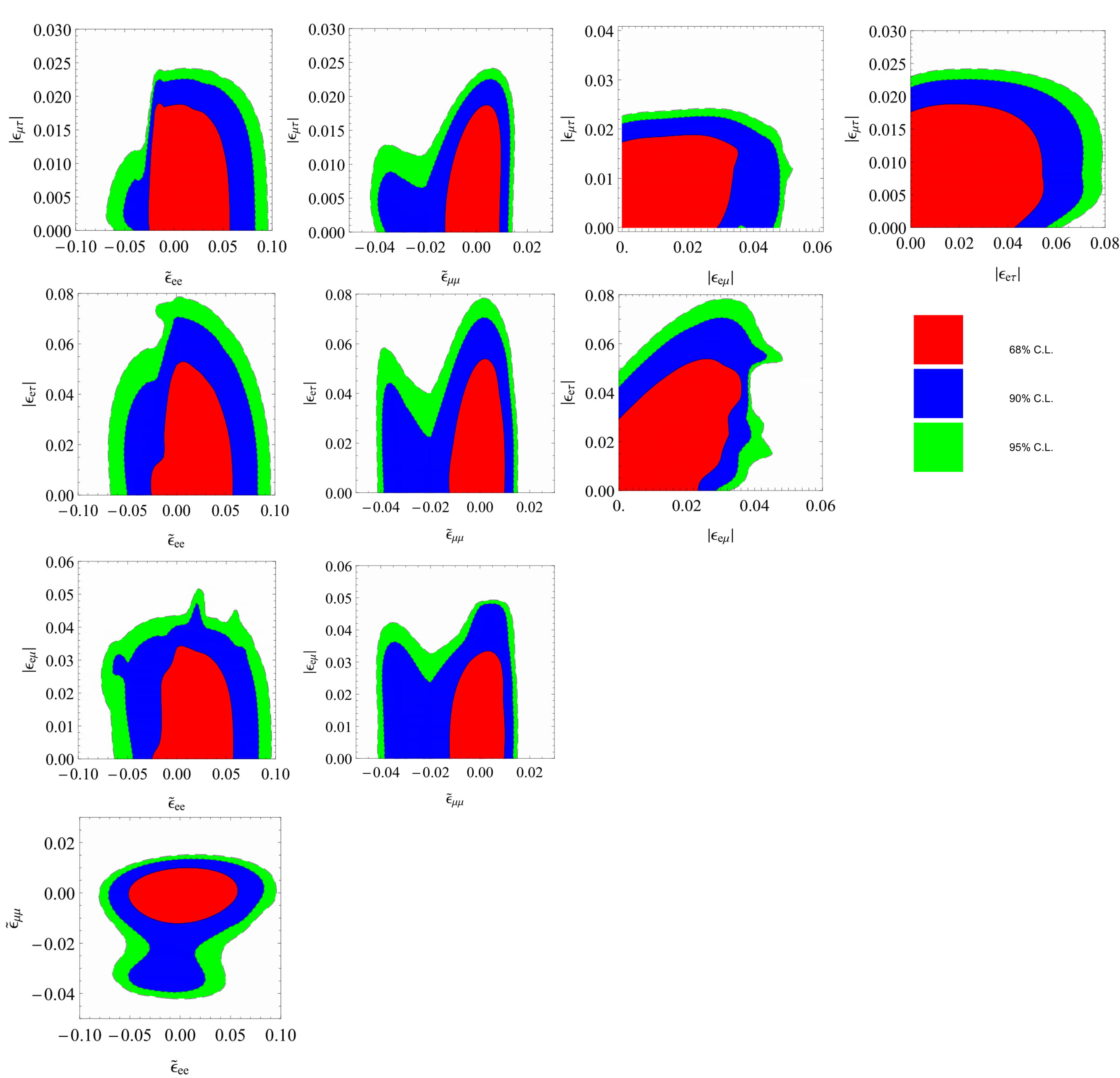} 
	\caption{The allowed regions on the plane spanned by two of NSI parameters. The central values and priors are taken from Table~\ref{tab:params}. All parameters are marginalized except those shown in each panel. }
	\label{fig:2D}
\end{figure*}

In Fig.~\ref{fig:2D}, we show the allowed region at~$68\%$, $90\%$, $95\%$ confidence level (C.L.) on the projected plan spanned by two of NSI parameters. 
{Though we do not see strong correlations in these results, a degeneracy for $\tilde{\epsilon}_{\mu\mu}\sim-0.035$ can be seen for contours of $90\%$ and $95\%$ C.L.
As mentioned in Sec.~\ref{sec:degeneracy}, this degeneracy is caused with the mixing angle $\theta_{23}$. Mentioned in Ref.~\cite{Coloma:2015kiu}, this degeneracy can be removed by including T2HK data.} Obviously, the degeneracy between $\tilde{\epsilon}_{ee}$ and $\epsilon_{\tau e}$ and the one around $\tilde{\epsilon}_{\mu\mu}=0.5$ for DUNE, are excluded.

\begin{figure*}[!bht]
	\centering
	\includegraphics[width=0.9\linewidth]{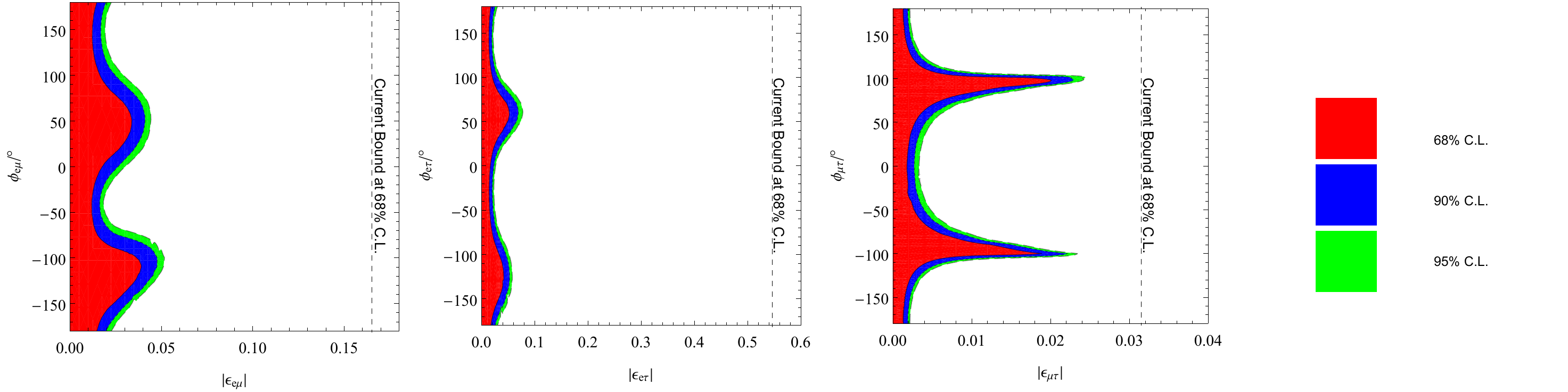}
	\caption{The allowed region on the plane spanned by the absolute value $|\epsilon_{\alpha\beta}|$ and the phase $\phi_{\alpha\beta}$ for off-diagonal elements
$\epsilon_{e\mu}$ (left), $\epsilon_{e\tau}$ (middle), and $\epsilon_{\mu\tau}$ (right) at 68\% (red), $90\%$ (blue) and $95\%$ (green) C.L..  All parameters are marginalized over except those shown in each panel.  }
	\label{fig:phase}
\end{figure*}

In Fig.~\ref{fig:phase}, we present allowed regions on the plane spanned by the absolute value $|\epsilon_{\alpha\beta}|$ and the phase $\phi_{\alpha\beta}$ for off-diagonal elements
$\epsilon_{e\mu}$, $\epsilon_{e\tau}$, and $\epsilon_{\mu\tau}$ at $68\%$, $90\%$ and $95\%$ C.L.. 
We see a strong correlation between the absolute value and the phase in these three panels. For the hypothesis with $\phi_{\alpha\beta}$ near $\pm90^\circ$, the bound on the absolute value $|\epsilon_{\alpha\beta}|$ is the worst. This behavior also applies to all off-diagonal elements. $\epsilon_{\alpha\beta}$ will be improved by more than an order of magnitude. Compared with sensitivities of NSIs at DUNE, our proposal will improve results by at least a factor of three. Of course, the phase of NSI parameters will play an important role here.

\subsection{CP violations of NSI}\label{sec:cpv}

We further discuss how the CERN-PINGU configuration can exclude the CP-conserved scenario where $\phi_{\alpha\beta}$ is neither $0$ nor $180^\circ$. We study the CPV sensitivity, which is defined
\begin{equation}\label{eq:NSI_CPV}
\begin{array}{l}
\Delta\chi^2_{CP}(\phi^{true}_{\alpha\beta})\equiv \min \{\Delta\chi^2(\phi_{\alpha\beta}=0),\\
\hspace{3.2cm}\Delta\chi^2(\phi_{\alpha\beta}=\pi)\},
\end{array}
\end{equation}
where $\phi^{true}_{\alpha\beta}$ is the true value for $\phi_{\alpha\beta}$, and $\Delta\chi^2(\phi_{\alpha\beta}=0)$ and $\Delta\chi^2(\phi_{\alpha\beta}=180^\circ)$ are the $\Delta\chi^2$ value for the hypothesis $\phi_{\alpha\beta}=0$ and $180^\circ$ respectively. This definition is same as how we study the sensitivity of CP violation where $\delta_\text{CP}$ is not $0$ or $180^\circ$. 

\begin{figure}[!ht]
	\centering
	\includegraphics[width=0.9\linewidth]{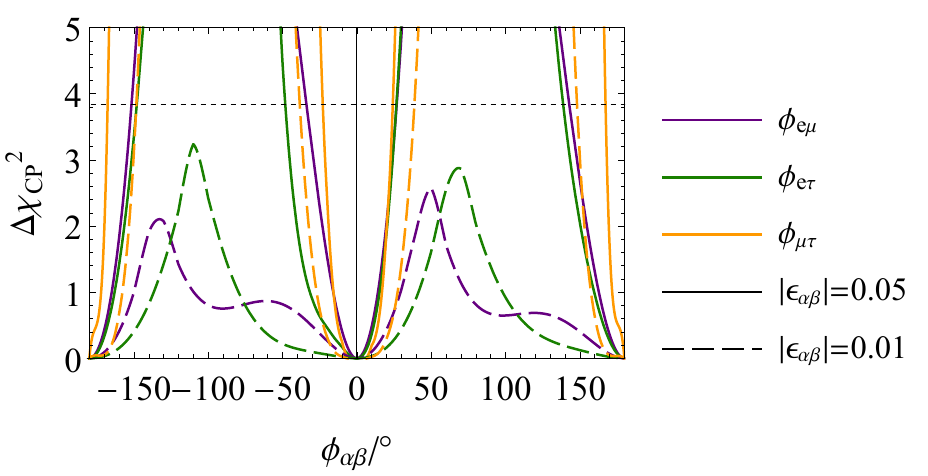}
	\caption{The $\Delta\chi^2_{CP}$ value against to the phase for the $e\mu$ (purple), $e\tau$ (green), and $\mu\tau$ (orange) components. For the studied component, the true absolute value is $0.05$ (solid) and $0.01$ (dashed), while the other NSIs are switched off.}
	\label{fig:CPV}
\end{figure}

In Fig.~\ref{fig:CPV}, we study the value of $\Delta\chi^2_{CP}$ for three phases: $\phi_{e\mu}$, $\phi_{e\tau}$ and $\phi_{\mu\tau}$ in the range between $-180^\circ$ and $180^\circ$. For $\Delta\chi^2_{CP}(\phi_{\alpha\beta})$, we fix the absolute value of the other two off-diagonal terms at $0$, as we focus on the CP violation by one specific $\phi_{\alpha\beta}$. Studying $\Delta\chi^2_{CP}(\phi_{\alpha\beta})$ values, we set the true value of $|\epsilon_{\alpha\beta}|$ to be $0.05$ (solid) and $0.01$ (dashed). 
We focus on the case with the absolute value fixed at $0.05$.
We see that the CPV sensitivity for $\phi_{\mu\tau}$ is the best. In about $81\%$ of possible $\phi_{\mu\tau}$  this configuration can exclude the CP-conserved scenario with the significance better than $95\%$ C.L.. $\Delta\chi^2_{CP}(\phi_{e\tau})$ is slightly better than $\Delta\chi^2_{CP}(\phi_{e\mu})$. For $\Delta\chi^2_{CP}(\phi_{e\tau})$ ($\Delta\chi^2_{CP}(\phi_{e\mu})$) larger than the value for $95\%$ C.L., it covers $75\%$ ($63\%$) of all possible phases. 
With the absolute value of $0.01$, the sensitivity is worse, especially for $\phi_{e\mu}$ and $\phi_{e\tau}$, of which $\Delta\chi^2_{CP}$ is smaller than $3.5$. For $\epsilon_{\mu\tau}$, phases with the sensitivity larger than $90\%$ C.L. cover about $75\%$ of all possible phases.
We note that the result shown in Fig.~\ref{fig:CPV} is consistent with what we see in Fig.~\ref{fig:prob_phase}. Our simulation result is based on the assumption that we focus on one specific off-diagonal element. Though not shown here, we find that once we also marginalize over absolute values and phases of the other off-diagonal terms, the sensitivity is much worse.

\subsection{Comparison with the other experiments}
\begin{table*}[!ht]
	\caption{The $1\sigma$ allowed range of $\tilde{\epsilon}_{ee}$, $\tilde{\epsilon}_{\mu\mu}$, $|\epsilon_{e\mu}|$, $|\epsilon_{e\tau}|$, and $|\epsilon_{\mu\tau}|$ for the CERN-PINGU configuration, DUNE, T2HK and P2O experiments. We also present the simulation details for each experiments.}
	\centering
	\begin{tabular}{c||c|c|c|c}
		\toprule
		parameter  
		&  CERN-PINGU & DUNE~\cite{Acciarri:2015uup} & T2HK~\cite{Hyper-Kamiokande:2016dsw} & P2O~\cite{Choubey:2019nlz,Akindinov:2019flp} \\
		\hline
		$ \widetilde{\epsilon}_{ee}$ & $\left[-0.0163, 0.0390\right]$ & $\left[-1.0025, 1.146\right]$ & $\left[-2.2069, 2.2934\right]$ & $\left[-0.1944, 0.1633\right]$ \\
		$ \widetilde{\epsilon}_{\mu\mu}$ & $\left[-0.0075, 0.0067\right]$ & $\left[-0.0753, 0.0769\right]$ & $\left[-0.2530, 0.2588\right]$ & $\left[-0.0409, 0.0328\right]$ \\
		$ |\epsilon_{e\mu}|$ & $\left[0, 0.0291\right]$ &  $\left[0, 0.1018\right]$ & $\left[0.4062\right]$ & $\left[0, 0.0346\right]$ \\
		$ |\epsilon_{e\tau}|$ & $\left[0, 0.0389\right]$ & $\left[0, 0.1938\right]$ & $\left[0, 0.2624\right]$ & $\left[0,0.039\right]$ \\
		$ |\epsilon_{\mu\tau}|$ & $\left[0, 0.0148\right]$ & $\left[0, 0.2012\right]$ & $\left[0, 0.9136\right]$ & $\left[0, 0.0719\right]$ \\ \hline\hline
		POTs & $10\times 10^{20}$ & $8.82\times10^{21}$ & $2.7\times 10^{22}$ & $2.4\times10^{20}$ \\\hline
		Energy range [GeV] & $3$-$20$ & $0.5$-$8$ & $0.1$-$1.2$ & $2$-$12$ \\\hline
		Baseline [km] & $11810$ & $1300$ & $295$ & $2595$ \\\hline 				
		Target material & ice & liquid argon & pure water & sea water \\\hline 
		Detector size [kton] & $\mathcal{O}(10^3)$ & $40$ & $186$ & $\mathcal{O}(10^3)$ \\\hline		
		\bottomrule
	\end{tabular}
	\label{tab:compare}
\end{table*}

In Table \ref{tab:compare}, we compare the CERN-PINGU configuration with the other future accelerator neutrino oscillation experiments, DUNE, T2HK, and the Protvino-ORCA experiment (P2O), with the simulation details about these configuration. For all of them, we use the PREM onion shell model of the earth for the matter density \cite{stacey_1977, Dziewonski:1981xy}, and include the prior that presents the constraint with the global and COHERENCE~\cite{Esteban:2018ppq}. Taking the same fluxes, effective masses, our $\nu$-mode simulation for P2O reproduces the event spectra in \cite{Choubey:2019nlz, Akindinov:2019flp}. However, because the spectra for $\bar{\nu}$ are not shown in this reference, we use the same detection efficiency for $\bar{\nu}$ same as those for $\nu$. The other assumption for detection is used the same as what we use for PINGU.
We find that the CERN-PINGU configuration performs the best among all, and that T2HK is the worst one to measure NSI parameters. The constraints of the CERN-PINGU configuration is smaller than DUNE by at least a factor of $0.1$. We need to point it out that for this result, DUNE requests0 more POTs than that for the CERN-PINGU configuration, by a factor of $\sim10$. 
The physics capability of P2O to measure NSI parameters, is the closet one to the CERN-PINGU configuration. The measurements of $|\epsilon_{e\mu}|$ and $|\epsilon_{e\tau}|$ for the CERN-PINGU configuration are slightly better than those for P2O. The constraint of $|\epsilon_{\mu\tau}|$ by the CERN-PINGU configuration can be $20\%$ of the result by P2O. Finally, the  measurement of the diagonal terms for the CERN-PINGU configuration can be improved by one order of magnitude, than that for P2O.

\section{Conclusion}\label{sec:conclusion}

With the interest on the planet-scale neutrino oscillations, we have considered the configuration that neutrinos are generated at CERN, and detected in the PINGU detector (CERN-PINGU), as an extended study of Ref.~\cite{Tang2012}. We have analyzed how such a configuration can measure the size of non-standard interactions (NSIs) in matter. Three advantages of this configuration are as follows: (1)~the energy range $3$~GeV-$20$~GeV enhancing effects of NSIs, (2)~the $11810$-km baseline cumulating these effects, and (3)~the high matter density $11$ g/cm$^3$ making the impact of NSIs on neutrino oscillations more significant.

We have adopted the GLoBES library with an extension package to simulate the CERN-PINGU configuration for the neutrino oscillation with NSI matter effects. We have studied the predicted uncertainty of NSI parameters $\epsilon_{\alpha\beta}$. We have found that this configuration measures $\tilde{\epsilon}_{\mu\mu}\equiv\epsilon_{\mu\mu}-\epsilon_{\tau\tau}$ and $|\epsilon_{\mu\tau}|$ better than the others.
Most of degeneracy problems for DUNE on $\epsilon_{\alpha\beta}$ measurements are resolved, except the one around $\tilde{\epsilon}_{\mu\mu}\sim-0.035$. 
We have investigated strong correlations between the absolute value and the phase for all off-diagonal terms, which can also be seen in the recent work with regards to NSI-parameter measurements for DUNE.
As phases $\phi_{\alpha\beta}$ play important roles, we have extended our study on the sensitivity of CP violation in NSIs. We have found that with the absolute value of $0.05$ the sensitivity can be better than $95\%$ C.L. in the CERN - PINGU configuration, which shows this configuration can be used to measure the CP violation caused by NSIs. {Compared to other experiments, DUNE, T2HK, and P2O, we find that the CERN-PINGU configuration can significantly improve the constraints of NSI parameters, except the measurements of $\epsilon_{e\mu}$ and $\epsilon_{e\tau}$, for which the CERN-PINGU configuration performs slightly better than P2O.}

This study should not be limited at the CERN-PINGU configuration. Our conclusion on the improvement of $\epsilon_{\alpha\beta}$ measurements could be applied for any experiments with the neutrino source or detector that satisfies three requirements: the proper energy range, a planet-scale baseline, and the high matter density. As PINGU is under consideration, we can put more focus on looking for possible sources, such as accelerator neutrinos produced by future proton drivers as super proton-proton collider (SPPC)~\cite{CEPC-SPPCStudyGroup:2015esa}, and the point source of astrophysical neutrinos~\cite{Aartsen:2019mbc}. 

\begin{acknowledgments}%TG
This work is supported in part by the grant National Natural Science Fundation of China under Grant No. 11505301 and No. 1188124024. JT appreciate ICTP's hospitality and discussions during the workshop PANE2018. Wei-Jie Feng and Yi-Xing Zhou are supported in part by Innovation Training Program for bachelor students at School of Physics in SYSU.
\end{acknowledgments}

\appendix
\begin{figure}[!ht]
	\centering
	\includegraphics[width=0.8\linewidth]{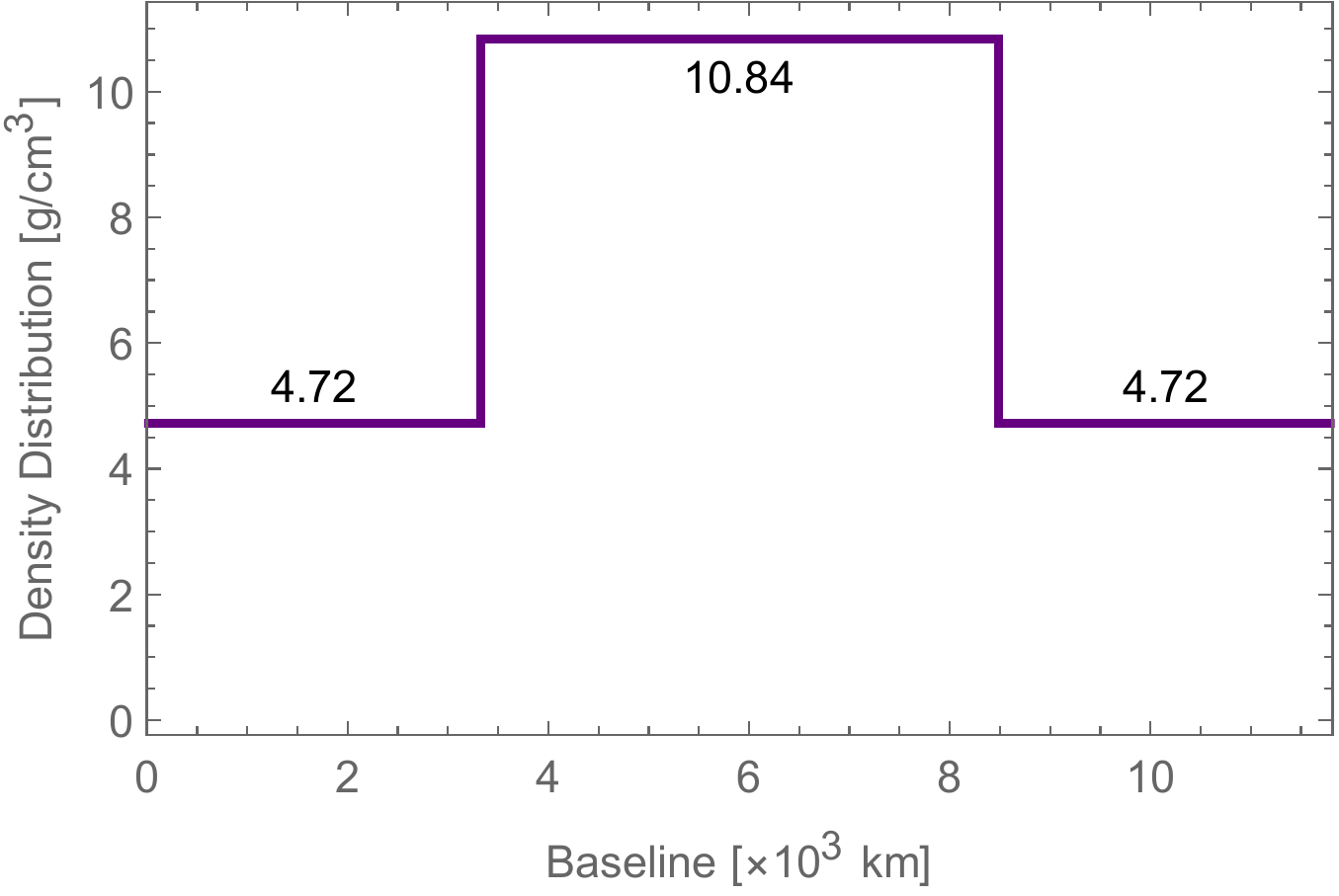}
	\caption{The matter density profile used in GLoBES simulation for the CERN-PINGU configuration. The x-axis is the distance from the source toward the detector.}
	\label{fig:density}
\end{figure}

\section{The matter density profile for CERN-PINGU}\label{App:matter_density}

Fig.~\ref{fig:density}, the matter density profile used in GLoBES simulation for the CERN-PINGU configuration. The x-axis is the distance from the source toward the detector. As we can see, the density can achieve up to $10.84$ g/cm$^3$ from the oscillation distance $3500$~km to $8000$~km.

\bibliographystyle{unsrt}
\bibliography{bibfile}% Produces the bibliography via BibTeX.

\end{document}